\documentclass[12pt]{article}
\usepackage{graphicx}
\begin{document}


\def\a{\alpha}
\def\b{\beta}
\def\c{\varepsilon}
\def\d{\delta}
\def\e{\epsilon}
\def\f{\phi}
\def\g{\gamma}
\def\h{\theta}
\def\k{\kappa}
\def\l{\lambda}
\def\m{\mu}
\def\n{\nu}
\def\p{\psi}
\def\q{\partial}
\def\r{\rho}
\def\s{\sigma}
\def\t{\tau}
\def\u{\upsilon}
\def\v{\B}
\def\w{\omega}
\def\x{\xi}
\def\y{\eta}
\def\z{\zeta}
\def\D{\Delta}
\def\G{\Gamma}
\def\H{\Theta}
\def\L{\Lambda}
\def\F{\Phi}
\def\P{\Psi}
\def\S{\Sigma}

\def\o{\over}
\newcommand{\gsim}{ \mathop{}_{\textstyle \sim}^{\textstyle >} }
\newcommand{\lsim}{ \mathop{}_{\textstyle \sim}^{\textstyle <} }
\newcommand{\vev}[1]{ \left\langle {#1} \right\rangle }
\newcommand{\bra}[1]{ \langle {#1} | }
\newcommand{\ket}[1]{ | {#1} \rangle }
\newcommand{\EV}{ {\rm eV} }
\newcommand{\KEV}{ {\rm keV} }
\newcommand{\MEV}{ {\rm MeV} }
\newcommand{\GEV}{ {\rm GeV} }
\newcommand{\TEV}{ {\rm TeV} }
\def\diag{\mathop{\rm diag}\nolimits}
\def\Spin{\mathop{\rm Spin}}
\def\SO{\mathop{\rm SO}}
\def\O{\mathop{\rm O}}
\def\SU{\mathop{\rm SU}}
\def\U{\mathop{\rm U}}
\def\Sp{\mathop{\rm Sp}}
\def\SL{\mathop{\rm SL}}
\def\tr{\mathop{\rm tr}}

\def\IJMP{Int.~J.~Mod.~Phys. }
\def\MPL{Mod.~Phys.~Lett. }
\def\NP{Nucl.~Phys. }
\def\PL{Phys.~Lett. }
\def\PR{Phys.~Rev. }
\def\PRL{Phys.~Rev.~Lett. }
\def\PTP{Prog.~Theor.~Phys. }
\def\ZP{Z.~Phys. }
\newcommand{\bear}{\begin{array}}  \newcommand{\eear}{\end{array}}
\newcommand{\bea}{\begin{eqnarray}}  \newcommand{\eea}{\end{eqnarray}}
\newcommand{\beq}{\begin{equation}}  \newcommand{\eeq}{\end{equation}}
\newcommand{\bef}{\begin{figure}}  \newcommand{\eef}{\end{figure}}
\newcommand{\bec}{\begin{center}}  \newcommand{\eec}{\end{center}}
\newcommand{\non}{\nonumber}  \newcommand{\eqn}[1]{\beq {#1}\eeq}
\newcommand{\lmk}{\left(}  \newcommand{\rmk}{\right)}
\newcommand{\lkk}{\left[}  \newcommand{\rkk}{\right]}
\newcommand{\lhk}{\left \{ }  \newcommand{\rhk}{\right \} }
\newcommand{\del}{\partial}  \newcommand{\abs}[1]{\vert{#1}\vert}
\newcommand{\vect}[1]{\mbox{\boldmath${#1}$}}
\newcommand{\bib}{\bibitem} \newcommand{\new}{\newblock}
\newcommand{\la}{\left\langle} \newcommand{\ra}{\right\rangle}
\newcommand{\bfx}{{\bf x}} \newcommand{\bfk}{{\bf k}}
\newcommand{\gtilde} {~ \raisebox{-1ex}{$\stackrel{\textstyle >}{\sim}$} ~} 
\newcommand{\ltilde} {~ \raisebox{-1ex}{$\stackrel{\textstyle <}{\sim}$} ~}
\newcommand{\gtrsim}{ \mathop{}_{\textstyle \sim}^{\textstyle >} }
\newcommand{\lesssim}{ \mathop{}_{\textstyle \sim}^{\textstyle <} }
\newcommand{\ds}{\displaystyle}
\newcommand{\bi}{\bibitem}
\newcommand{\lar}{\leftarrow}
\newcommand{\rar}{\rightarrow}
\newcommand{\lrar}{\leftrightarrow}
\def\Frac#1#2{{\displaystyle\frac{#1}{#2}}}
\def\labelenumi{(\roman{enumi})}
\def\SEC#1{Sec.~\ref{#1}}
\def\FIG#1{Fig.~\ref{#1}}
\def\EQ#1{Eq.~(\ref{#1})}
\def\EQS#1{Eqs.~(\ref{#1})}
\def\lrf#1#2{ \left(\frac{#1}{#2}\right)}
\def\lrfp#1#2#3{ \left(\frac{#1}{#2}\right)^{#3}}
\def\GEV#1{10^{#1}{\rm\,GeV}}
\def\MEV#1{10^{#1}{\rm\,MeV}}
\def\KEV#1{10^{#1}{\rm\,keV}}


\baselineskip 0.7cm

\begin{titlepage}

\begin{flushright}
\hfill DESY 06-106\\
\hfill hep-ph/0607170\\
\hfill November, 2006\\
\end{flushright}

\vskip 1.35cm
\begin{center}
{\large \bf
Inflaton Decay through Supergravity Effects
}
\vskip 1.2cm
Motoi Endo$^{1,2}$, Masahiro Kawasaki$^{1}$, Fuminobu Takahashi$^{1,2}$ 
and T. T. Yanagida$^{3,4}$
\vskip 0.4cm

${}^1${\it Institute for Cosmic Ray Research,
     University of Tokyo, \\ Chiba 277-8582, Japan}\\
${}^2${\it Deutsches Elektronen Synchrotron DESY, Notkestrasse 85,\\
22607 Hamburg, Germany}\\
${}^3${\it Department of Physics, University of Tokyo,\\
     Tokyo 113-0033, Japan}\\
 ${}^4${\it Research Center for the Early Universe, University of Tokyo,\\
     Tokyo 113-0033, Japan}

\vskip 1.5cm

\abstract{ We point out that supergravity effects enable the inflaton
to decay into all matter fields, including the visible and the
supersymmetry breaking sectors, once the inflaton acquires a
non-vanishing vacuum expectation value. The new decay processes have
great impacts on cosmology; the reheating temperature is bounded
below; the gravitinos are produced by the inflaton decay in a broad
class of the dynamical supersymmetry breaking models.  We derive the
bounds on the inflaton mass and the vacuum expectation value, which
severely constrain high-scale inflations such as the hybrid and
chaotic inflation models. }
\end{center}
\end{titlepage}

\setcounter{page}{2}

\section{Introduction}
\label{sec:1}

In recent articles~\cite{Kawasaki:2006gs}, we pointed out that
high-scale inflations such as hybrid inflation
models~\cite{Copeland:1994vg} are disfavored, since too many
gravitinos are produced in reheating processes after the inflation if
the hidden-sector field $z$ for supersymmetry (SUSY) breaking is
completely neutral~\cite{moduli,Asaka:2006bv,Dine:2006ii,Endo:2006tf}.
However if the $z$ has charges of some symmetries like in gauge- and
anomaly-mediation models~\cite{GMSB,AMSB}, dangerous operators are
suppressed~\cite{Dine:2006ii, Endo:2006tf,Kawasaki:2006gs} and there
is no gravitino-overproduction problem~\footnote{See however
Ref.~\cite{Endo:2006tf} for other potential problems due to
non-renormalizable couplings between the inflaton and the $z$.}.
Furthermore, such symmetries may suppress linear terms of the field
$z$ in K\"ahler potential to avoid the Polonyi
problem~\cite{Ibe:2006am}.

In this letter, we show that, as long as kinematically allowed, the
inflaton $\phi$ decays into all matter fields that appear in the
superpotential due to the supergravity (SUGRA) effect, once the
inflaton acquires a non-vanishing vacuum expectation value (VEV), $\la
\phi \ra$.  For a typical inflaton mass in high-scale inflations, the
inflaton may decay into the SUSY breaking sector fields, producing the
$z$ field and the gravitinos in a broad class of the dynamical SUSY
breaking (DSB) models~\cite{Witten:1981nf,Affleck:1984uz, Dine:1993yw,
dns,Izawa:1996pk,Intriligator:1996pu}.  Since the decay proceeds
independently of the charge of the $z$, the gravitino production from
the inflaton decay is a generic problem, which is present even in the
gauge- and anomaly-mediation models.  In particular, the gravitinos
are produced even if the K\"ahler potential is minimal. Further, the
inflaton with a nonzero VEV can decay into the visible sector fields
through the top Yukawa coupling, even if there is no direct
interaction in the global SUSY limit.  This enables us to set a lower
limit on the reheating temperature of the inflation models. We derive
constraints on the inflaton mass and the VEV, taking account of those
new decay processes.

\section{Decay Processes}
\label{sec:2}

The inflaton field $\phi$ couples to all matter fields due to the
SUGRA effects, assuming a non-vanishing VEV.  The relevant
interactions with fermions are represented in terms of the total
K\"ahler potential $G = K + \ln |W|^2$ in the Planck unit $M_P = 1$:
\begin{eqnarray}
  \mathcal{L} \;=\; 
  - \frac{1}{2} e^{G/2} G_{\phi ijk}\,
  \phi\,\psi^i\psi^j\varphi^k + h.c.,
\end{eqnarray}
where $\varphi^i$ denotes a scalar field and $\psi^i$ is a fermion in
2-spinor representation. We assume $G_i \ll O(1)$ for all the fields
other than the SUSY breaking field. The contribution proportional to
$G_\phi$ may be suppressed in the inflaton decay because of the
interference with the SUSY breaking
field~\cite{Dine:2006ii,Endo:2006tf}.

In this section, we assume the minimal K\"ahler potential for
simplicity. Then, there is no non-renormalizable term in the K\"ahler
potential. Even when the inflaton field is secluded from the other
fields in the global SUSY Lagrangian, the SUGRA corrections make its
decay possible. At the vacuum, the coupling constants are expanded as
\begin{eqnarray}
  G_{\phi ijk} \;=\; 
  -\frac{W_\phi}{W} \frac{W_{ijk}}{W} + \frac{W_{\phi ijk}}{W} 
  \;\simeq\; K_\phi \frac{W_{ijk}}{W} + \frac{W_{\phi ijk}}{W},
\end{eqnarray}
where we assumed that the VEVs are negligibly small for all the fields
other than the inflaton, and used $G_\phi \ll \la \phi \ra$ in the
last equality.  We find that the result is obviously invariant under
the K\"ahler transformation, and these constants approach to zero in
the global SUSY limit.  Then the decay rates are evaluated
as~\footnote{
  Similarly we obtain the 2-body decays, 
  though the decay rate is
  suppressed by $O(10^2) \times M_{ij}^2/m_\phi^2$ with $M_{ij} \equiv
  W_{ij}$ compared to the 3-body decay rate with $Y = O(1)$. 
  See also Ref.~\cite{Allahverdi:2001ux}. 

}
\begin{eqnarray}
  \Gamma(\phi \rightarrow \psi^i\psi^j\varphi^k) 
  &\simeq&
  \frac{N_f}{1536\pi^3} 
  |Y_{ijk}|^2 
  \lrfp{\langle \phi \rangle}{M_P} {2}
  \frac{m_\phi^3}{M_P^2},
\end{eqnarray}
where $N_f$ is a number of the final state, and the Yukawa coupling
$Y_{ijk} \equiv W_{ijk}$. Here we have neglected the masses of the
final state particles, and used $K = \phi^\dagger\phi$ for the
inflaton. We have also assumed that $\psi^i$ and $\psi^j$ are not
identical particles. It is stressed again that the inflaton decay
proceeds even when there is no direct interactions with the matter
fields in the global SUSY limit.

The decay rates of the inflaton into the scalar particles become the
same as the above results. In fact, with the scalar potential, $V =
e^G(G^i G_i - 3)$, the decay amplitude of $\phi^* \rightarrow
\varphi^i \varphi^j \varphi^k$ is estimated as $V_{\bar \phi
ijk}$. Since the inflaton has a large SUSY mass, the amplitude is
approximately proportional to $e^{G/2} G_{\phi ijk}$ multiplied by the
inflaton mass, $m_\phi \simeq e^{G/2}
|G^\phi_{\phantom{a}\bar\phi}|$. Therefore the decay rate satisfies
$\Gamma(\phi^* \rightarrow \varphi^i \varphi^j \varphi^k) \simeq 3
\Gamma(\phi \rightarrow \psi^i \psi^j \varphi^k)$.

\subsection{Lower bound on the reheating temperature}

Let us consider the inflaton decay through the top Yukawa coupling:
\beq
W \;=\; Y_t \,T Q H_u,
\eeq
where $Y_t$ is the top Yukawa coupling, and $T$, $Q$, and $H_u$ are
the chiral supermultiplets of the right-handed top quark and
left-handed quark doublet of the third generation, and up-type Higgs,
respectively. The partial decay rate of the inflaton through the top
Yukawa coupling is
\beq
\label{eq:rate-th-top}
\Gamma_T \;=\; \frac{3}{128 \pi^3} |Y_t|^2 \lrfp{\la \phi \ra}{M_P}{2} 
\frac{m_\phi^3}{M_P^2},
\eeq
which sets a lower bound on the reheating temperature, $T_R$.  We
define the reheating temperature as
\beq
\label{eq:def-Tr}
T_R \;\equiv\; \lrfp{\pi^2 g_*}{10}{-\frac{1}{4}} \sqrt{\Gamma_\phi M_P},
\eeq
where $g_* $ counts the relativistic degrees of freedom, and
$\Gamma_\phi$ denotes the total decay rate of the inflaton.  Using
Eqs.~(\ref{eq:rate-th-top}) and (\ref{eq:def-Tr}), we obtain the lower
bound on the reheating temperature,
\beq
T_R \;\gtrsim\; 1.9 \times 10^3 {\rm\,GeV}\, |Y_t| 
\lrfp{g_*}{200}{-\frac{1}{4}} 
\lrf{\la \phi \ra}{10^{15}{\rm \,GeV}}
\lrfp{m_\phi}{\GEV{12}}{\frac{3}{2}}.
\eeq
We show the contours of the lower limit on the reheating temperature
in Fig.~\ref{fig:tr-low}, together with the results of the hybrid and
smooth hybrid inflation models~\cite{Copeland:1994vg,
Lazarides:1995vr}.  If the inflaton mass $m_\phi$ and the VEV $\la
\phi \ra$ are too large, the reheating temperature may exceed the
upper bound due to the abundance of the gravitinos produced by
particle scattering in the thermal plasma; the recent results are
given in~\cite{Kawasaki:2004yh, Kohri:2005wn, Jedamzik:2006xz} for the
unstable gravitino and in~\cite{Moroi:1993mb} for the stable one (see
Ref.~\cite{Kawasaki:2006gs} for the summarized results).  In
particular, for the smooth hybrid inflation model, the reheating
temperature is necessarily higher than $10^6{\rm\,GeV}$, which is
difficult to be reconciled with the gravitino of $m_{3/2} \;=\;
O(0.1-1){\rm\,TeV}$~\cite{Kawasaki:2004yh}.

\subsection{Inflaton decay into the gravitinos}

The new decay process enables the inflaton to decay into the SUSY
breaking sector. Through Yukawa interactions containing the SUSY
breaking field, the gravitino is produced from the inflaton decay.  It
should be noted that the gravitino can be produced even without
non-renormalizable coupling $|\phi|^2 z z$ in the K\"ahler potential.
The existence of such coupling is crucial for the gravitino pair
production~\cite{Kawasaki:2006gs,Endo:2006tf}, if the inflaton mass is
larger than the scalar mass of the SUSY breaking field $z$.

Let us consider the Yukawa interactions of the SUSY breaking sector
fields~\footnote{
  If there exists a linear term $W = \mu^2 z$ in the superpotential,
  the inflaton $\phi$ mixes with $z$ in the vacuum.  Even if the
  minimum of $z$ during inflation coincides with that after inflation,
  the coherent oscillations of $z$ field is induced by the inflaton
  via the mixing~\cite{Endo:2006tf}.  However, the induced amplitude
  of $z$ is so small that it is cosmologically harmless.
}.  From the point of view of naturalness, the DSB
scenarios~\cite{Witten:1981nf} are attractive. In a wide class of the
DSB models~\cite{Affleck:1984uz, Dine:1993yw, dns,Izawa:1996pk,
Intriligator:1996pu}, including ones reduced from ${\cal N}=2$ SUSY
quiver gauge theories~\cite{Ooguri:2006pj}, there exist such Yukawa
interactions at the quark level~\footnote{
  Note, however, that those models without Yukawa interactions can
  also break the SUSY dynamically. Examples include SU(5) model with
  ${\bf 5}^*$ and ${\bf 10}$, and SO(10) with ${\bf
  16}$~\cite{Affleck:1983vc,Affleck:1984uz}. Then the constraint due
  to the gravitino production discussed in this letter is not applied
  for such models.
}.  When one considers the inflaton decay into the SUSY breaking
sector, the quark-level interactions should be used, instead of the
one written in terms of the composite fields in the low-energy
effective theory, since the inflaton mass scale is larger than the
strong-coupling scale. The fields in the SUSY breaking sector
typically acquire masses of the DSB scale $\Lambda \sim \sqrt{m_{3/2}
M_P}$, while the inflaton in the high-scale inflation models such as
the hybrid inflation model tends to satisfy with $m_\phi \gtrsim
\Lambda$. Then the inflaton can decay into the hidden sector fields.

The SUSY breaking field in the low-energy effective theory is either
an elementary or composite field. Let us denote the SUSY breaking
field by $z$ throughout this letter. If the $z$ is an elementary field
and appears in a Yukawa interaction, its fermionic component is
directly produced by the inflaton decay. On the other hand, if $z$ is
composite, the produced hidden quarks will form QCD-like jets, ending
up with the hidden mesons and/or baryons, together with (possibly
many) $z$ fields.  In both cases, since the fermionic component of $z$
is identified with the goldstino, the gravitino is produced by the
inflaton decay.  In addition, the scalar component of $z$ and the
other SUSY breaking sector fields may also decay into the gravitino.

Therefore the gravitino production rate is expressed as~\footnote{
  The same result can be obtained from the SUGRA Lagrangian.
  Actually, assuming the Yukawa coupling $W = zQQ$, the decay rate of
  $\phi \rightarrow \tilde G QQ$ in the goldstino picture is
  reproduced by evaluating the $Q$-mediating diagram in the unitary
  gauge, where $\tilde G$ is the goldstino.
}
\beq
\label{gravitino-rate}
\Gamma_{3/2} \;=\; 
\frac{C}{1536 \pi^3}
\lrfp{\la \phi \ra}{M_P}{2}
\frac{m_\phi^3}{M_P^2},
\eeq
where $C$ is determined by the decay processes; the degrees of freedom
of the decay products, the decay chains, the coupling constants of the
Yukawa interactions in the SUSY breaking sector, and form factors of
the hidden mesons and/or baryons. Although the constant $C$ strongly
depends on the models, unless all the Yukawa couplings are extremely
suppressed, $C$ is expected to be $O(1)$ or larger.  The gravitino
abundance is then given by~\footnote{ We have assumed here that there
is no entropy production~\cite{Lyth:1995ka} after the reheating.  }
\beq
Y_{3/2} \;=\; 
8 \times 10^{-14}\, C 
\lrfp{g_*}{200}{-\frac{1}{2}} 
\lrfp{T_R}{10^6{\rm\,GeV}}{-1}
\lrfp{m_\phi}{10^{12}{\rm\,GeV}}{2} 
\lrfp{\la\phi\ra}{10^{15}{\rm\,GeV}}{2},
\label{gravitinoY}
\eeq
for $m_\phi \gtrsim \Lambda$.

For the inflaton mass, $m_\phi \lesssim \Lambda$, the inflaton decay
into the SUSY breaking sector is likely to be kinematically
forbidden. However, the gravitino pair production from the inflaton
decay instead becomes
important~\cite{Kawasaki:2006gs,Dine:2006ii,Endo:2006tf}, if the
inflaton mass is smaller than the mass of the scalar component of the
$z$ field.  We here assume that the mass of the $z$ field is
$O(\Lambda)$ in the DSB scenarios.  The gravitino pair production rate
is given by~\cite{Kawasaki:2006gs,moduli,Endo:2006tf}
\beq
\Gamma_{3/2}^{\rm(pair)} \;=\; \frac{3}{288 \pi} 
\lrfp{\la \phi \ra}{M_P}{2} \frac{m_\phi^3}{M_P^2},
\eeq
which is larger than Eq.~(\ref{gravitino-rate}) by two orders of
magnitude. Note that, as long as $m_\phi \lesssim \Lambda$, the
inflaton decay into a pair of the gravitinos occurs at the rate shown
above, even if the K\"ahler potential is minimal.  The gravitino
abundance is then given by
\beq
Y_{3/2} \;=\; 
2 \times 10^{-11}\, 
\lrfp{g_*}{200}{-\frac{1}{2}} 
\lrfp{T_R}{10^6{\rm\,GeV}}{-1}
\lrfp{m_\phi}{10^{12}{\rm\,GeV}}{2} 
\lrfp{\la\phi\ra}{10^{15}{\rm\,GeV}}{2},
\label{gravitinoY-pair}
\eeq
for $m_\phi \lesssim  \Lambda$.

There are tight constraints on the gravitino abundance from BBN if the
gravitino is unstable, and from the dark matter (DM) abundance for the
stable gravitino.  In particular, for the gravitinos produced by
thermal scatterings, the gravitino abundance is related to $T_R$ by
using~\cite{Bolz:2000fu,Kawasaki:2004yh}
\begin{eqnarray}
    \label{eq:Yx-new}
    Y_{3/2}^{(th)} &\simeq& 
    1.9 \times 10^{-12}\left[ 1+ 
    \left(\frac{m_{\tilde{g}_3}^2}{3m_{3/2}^2}\right)\right]
    \left( \frac{T_{\rm R}}{10^{10}\ {\rm GeV}} \right)
    \nonumber \\ 
    & \times & 
    \left[ 1 
        + 0.045 \ln \left( \frac{T_{\rm R}}{10^{10}\ {\rm GeV}} 
        \right) \right]
    \left[ 1 
        - 0.028 \ln \left( \frac{T_{\rm R}}{10^{10}\ {\rm GeV}} ,
        \right) \right],
\end{eqnarray}
where we have taken $N=3$ for QCD and $m_{\tilde{g}_3}$ is the gluino
mass evaluated at $T=T_R$.  Since the gravitino abundance
$Y_{3/2}^{(th)}$ is roughly proportional to $T_R$, both are bounded
above. Substituting the upper bounds into Eqs.~(\ref{gravitinoY}) and
(\ref{gravitinoY-pair}), we obtain constraints on the mass and the VEV
of the inflaton.

Here we simply quote the bounds on $Y_{3/2}$ and $T_R$ summarized in
Ref.~\cite{Kawasaki:2006gs}, for representative values of the
gravitino mass: $m_{3/2} = 1{\rm\,GeV}$, $1{\rm\,TeV}$ and
$100{\rm\,TeV}$.  For the stable gravitino, the gravitino abundance
should not exceed the DM abundance~\cite{Moroi:1993mb},
\beq
m_{3/2} \,Y_{3/2} \;\leq\; \Omega_{\rm DM} \frac{\rho_c}{s} \;\lesssim\;
4.7 \times 10^{-10} {\rm\,GeV},
\label{eq:gra-abu}
\eeq
where $\rho_c$ is the critical density, and we used $\Omega_{\rm DM}
h^2 \lesssim 0.13$ at $95\%$ C.L.~\cite{Spergel:2006hy} in the second
inequality. Here we have neglected the contribution from the decay of
the next-to-lightest SUSY particle. The upper bound on $T_R$ can be
obtained by substituting \EQ{eq:gra-abu} into \EQ{eq:Yx-new} as
\begin{equation}
\label{eq:stable-g}
    T_R ~\lesssim ~3\times 10^7~{\rm GeV} 
   \lrfp{m_{\tilde{g}_3}}{500{\rm\,GeV}}{-2}   
\left(\frac{m_{3/2}}{1{\rm \,GeV}}\right),
\end{equation}
for $m_{3/2}\simeq 10^{-4} - 10~{\rm GeV}$.  For the unstable
gravitinos with $m_{3/2} = 1{\rm\,TeV}$, the bounds come from
BBN:~\cite{Kawasaki:2006gs}
\begin{eqnarray}
\label{eq:unstable-Y1}
   Y_{3/2}  & ~ \lesssim &\left\{
 \bear{cc}   4\times 10^{-17}, &~~~(B_h \simeq 1),\\
		 3 \times 10^{-14}, &~~~(B_h \simeq 10^{-3}),
 \eear		      
 \right.
\end{eqnarray}
where $B_h$ denotes the hadronic branching ratio of the gravitino. The
bounds on $T_R$ are given by
\begin{eqnarray}
\label{eq:unstable-T1}
   T_R & ~ \lesssim & 
   \left\{
 \bear{cc}   3 \times 10^{5}{\rm\,GeV}, &~~~(B_h \simeq 1),\\
		 2 \times 10^{8}{\rm\,GeV}, &~~~(B_h \simeq 10^{-3}),
 \eear		      
 \right.
\end{eqnarray}
For the unstable gravitino with $m_{3/2} = 100{\rm\,TeV}$, the
constraint comes from the abundance of the lightest SUSY particle
(LSP) produced by the gravitino decay.  In the anomaly-mediated SUSY
breaking models with the wino LSP, one LSP remains as a result of the
decay of one gravitino, since the gravitino decay temperature is
rather low. The bounds read
\bea
Y_{3/2} & \lesssim & 2 \times 10^{-12} 
\lrfp{m_{3/2}}{100{\rm\,TeV}}{-1} ,\non\\
T_R & \lesssim & 9 \times 10^{9} 
\lrfp{m_{3/2}}{100{\rm\,TeV}}{-1} {\rm GeV},
\label{eq:const-from-winolsp}
\eea
where we have used the relation,
\begin{equation}
  \label{eq:winomass}
  m_{\tilde{W}} \;=\; \frac{\beta_2}{g_2} \,m_{3/2} 
  \simeq 2.7 \times 10^{-3} m_{3/2},
\end{equation}
where $\beta_2$ and $g_2$ are the beta function and the gauge coupling
of $SU(2)_L$.

\begin{figure}[t]
\begin{center}
\includegraphics[width=10cm]{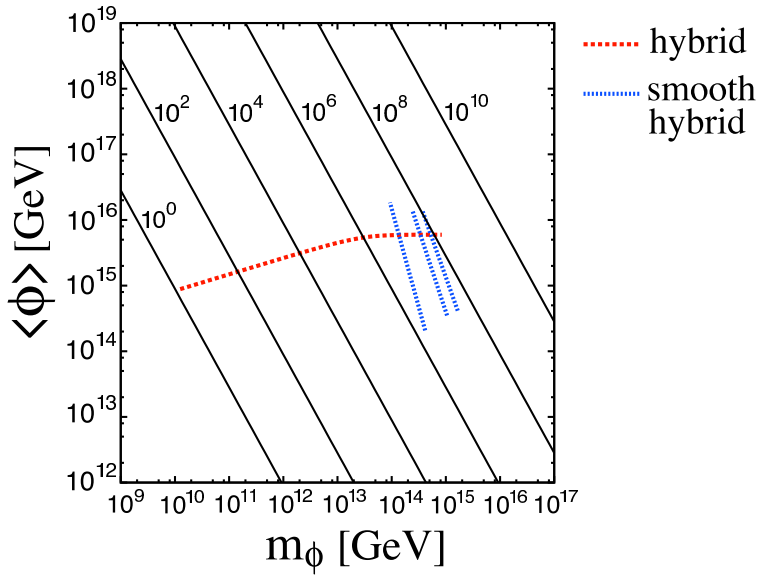}
\caption{Contours of the lower bound on the reheating temperature
$T_R$, denoted by the solid lines. We set $g_* \;=\; 228.75$ and $Y_t
\;=\; 0.6$.  The typical values of $\la \phi \ra$ and $m_\phi$ for the
hybrid and smooth hybrid inflation models ($n= 2, \,3$ and $4$ from
left to right) are also shown. }
\label{fig:tr-low}
\end{center}
\end{figure}

\begin{figure}[t]
\begin{center}
\includegraphics[width=10cm]{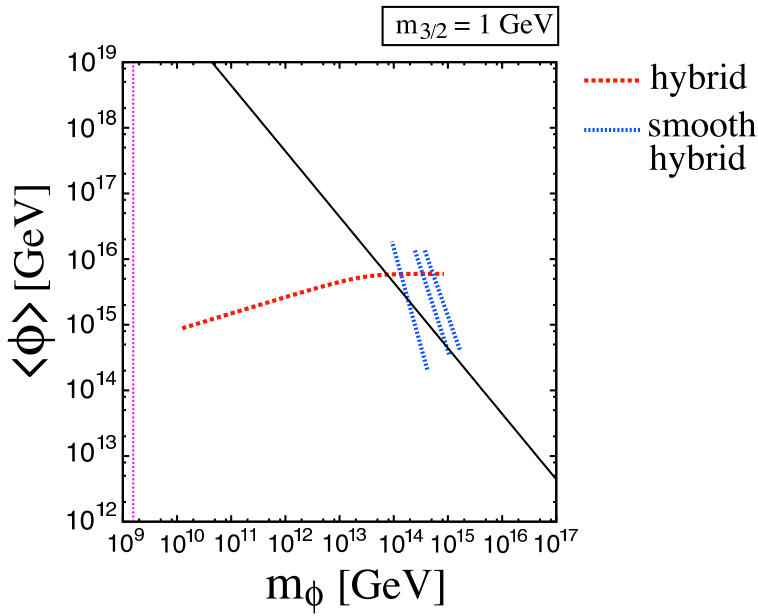}
\caption{Constraints from the abundance of the gravitino produced from
the inflaton decay, for $m_{3/2} = 1{\rm\,GeV}$. We set $g_* \;=\;
228.75$ and $C \;=\; 1$. The region above the solid line is excluded.
$T_R$ is set to be the largest allowed value, and the bound becomes
severer for lower $T_R$. The vertical dotted line corresponds to
$m_\phi = \sqrt{m_{3/2} M_P}$. For $m_\phi \lesssim \sqrt{m_{3/2}
M_P}$, the inflaton decay into the SUSY breaking sector is expected to
be kinematically forbidden; however the gravitino pair production
instead puts a severer
constraint~\cite{moduli,Asaka:2006bv,Dine:2006ii,Endo:2006tf}.  The
typical values of $\la \phi \ra$ and $m_\phi$ for the hybrid and
smooth hybrid inflation models are also shown. }
\label{fig:1GeV}
\end{center}
\end{figure}

\begin{figure}[t]
\begin{center}
\includegraphics[width=10cm]{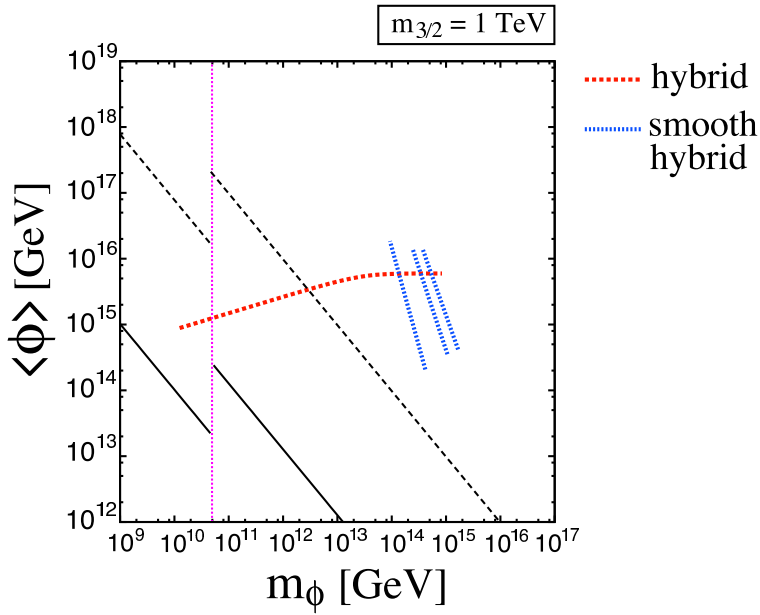}
\caption{Same as Fig.~\ref{fig:1GeV} but for $m_{3/2} = 1{\rm\,TeV}$.
The solid and dashed lines are for the hadronic branching ratio $B_h =
1$ and $10^{-3}$, respectively.  }
\label{fig:1TeV}
\end{center}
\end{figure}

\begin{figure}[t]
\begin{center}
\includegraphics[width=10cm]{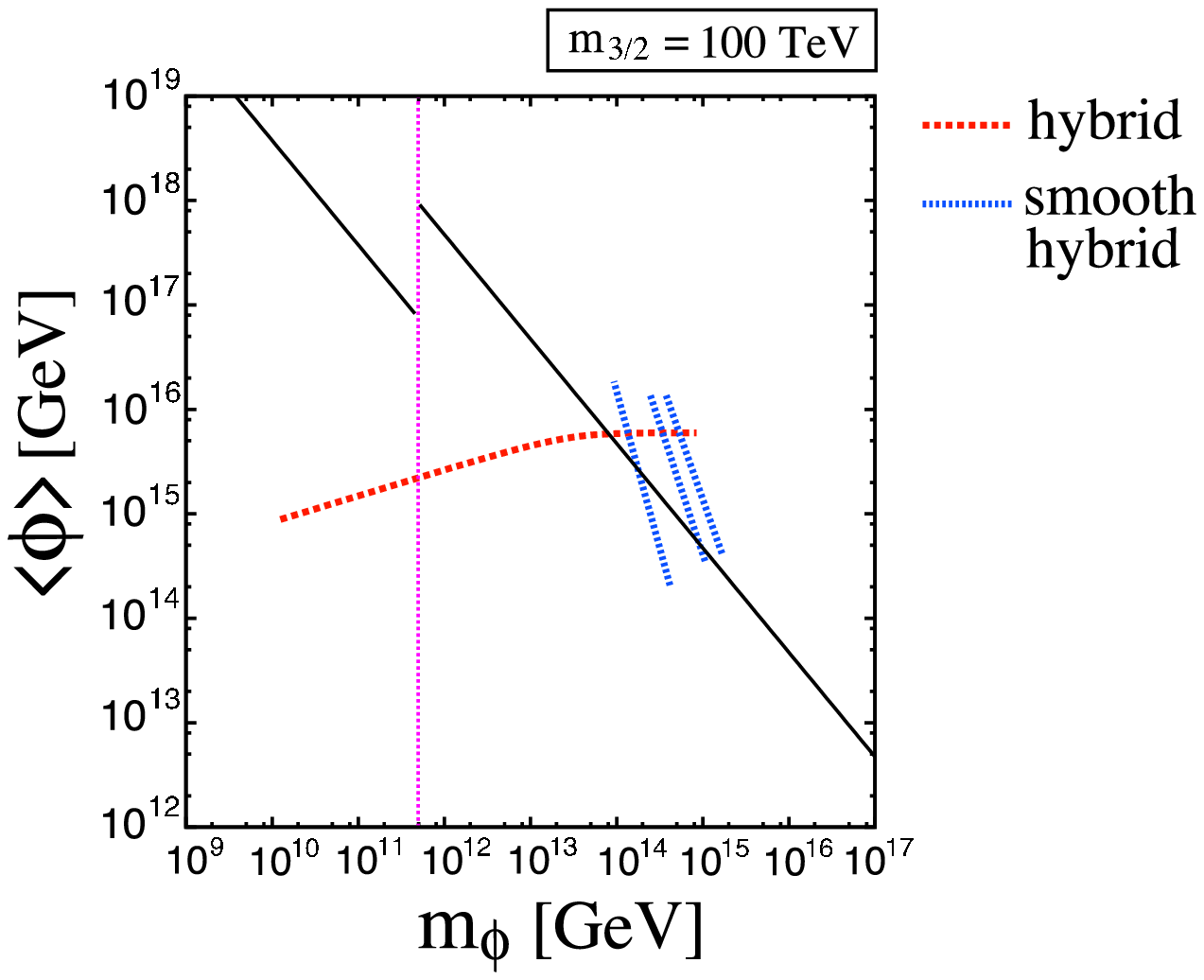}
\caption{Same as Fig.~\ref{fig:1GeV} but for $m_{3/2} =
100{\rm\,TeV}$. }
\label{fig:100TeV}
\end{center}
\end{figure}

We show the bounds on the mass and the VEV of the inflaton, obtained
by substituting the above upper bounds on $Y_{3/2}$ and $T_R$ into
Eqs.~(\ref{gravitinoY}) and (\ref{gravitinoY-pair}), in
Figs.~\ref{fig:1GeV}, \ref{fig:1TeV} and \ref{fig:100TeV}.  The bounds
are severer than that due to the lower bound on $T_R$ shown in
Fig.~\ref{fig:tr-low}, and a significant fraction of the parameter
space is excluded by the gravitino production. The bound shown in
Fig.~\ref{fig:1GeV} does not change for the stable gravitinos with
$m_{3/2} \simeq 100{\rm\,keV} - 10{\rm\,GeV}$, since both the upper
bound on $T_R$ is proportional to $m_{3/2}$.  It should be noted that
the bounds become severer for lower $T_R$, since $T_R$ is set to be
the largest allowed value.

\section{Constraints on Inflation Models}
\label{sec:3}
To see the impacts of the decay processes discussed in the previous
section, let us first consider the hybrid inflation
model~\cite{Copeland:1994vg}.  The hybrid inflation model contains two
kinds of superfields: one is $\phi$ which plays a role of inflaton and
the others are waterfall fields $\psi$ and $\tilde{\psi}$.

The superpotential $W(\phi, \psi,\tilde{\psi})$ for the hybrid inflaton is
\begin{equation}
   \label{eq:spot_hyb}
   W(\phi, \psi,\tilde{\psi}) \;=\; \phi (\mu^{2}  
   - \lambda \tilde{\psi}\psi),
\end{equation}
where $\mu$ determines the inflation energy scale, and $\psi$ and
$\tilde \psi$ are assumed to be charged under $U(1)$ gauge symmetry.
Here $\lambda$ is a coupling constant and $\mu$ is the inflation
energy scale. The potential minimum is located at $\la\phi\ra = 0$ and
$\langle \psi\rangle = \langle\tilde{\psi}\rangle =
\mu/\sqrt{\lambda}$ in the SUSY limit.  For a successful
inflation~\footnote{ We require the right magnitude of the density
fluctuations and the spectral index less than or equal to
unity~\cite{Spergel:2006hy}. The e-folding number is set to be $N_e =
50$.  }, $\mu$ and $\lambda$ are related as
\bea
\mu &\simeq& 2\times 10^{-3}\,\lambda^{1/2} 
\rm{~~~ for~~~} \lambda \;\gtrsim \;10^{-3}, \non\\
\mu &\simeq & 2\times 10^{-2}\,\lambda^{5/6} 
\rm{~~~ for~~~} \lambda \;\lesssim\; 10^{-3},
\eea 
where $\lambda$ varies from $10^{-5}$ to 
$10^{-1}$~\cite{Bastero-Gil:2006cm}~\footnote{
Note that, in this type of hybrid inflation, there exists a problem of
cosmic string formation because $\psi$ and $\tilde\psi$ have $U(1)$
gauge charges. To avoid the problem the coupling $\lambda$ should be
small as, $\lambda \lesssim 10^{-4}$~\cite{Endo:2003fr}. Here we do
not take this problem seriously, since the cosmic strings are not
produced if the gauge group is extended to a non-Abelian
group~\cite{Dvali:1994ms,Watari:2004xh}.
}.

After inflation ends, the universe is dominated by both the inflaton
$\phi$ and a combination of the waterfall fields, $\psi_{+} \equiv
(\psi + \tilde \psi)/\sqrt{2}$, while the other combination, $\psi_{-}
\equiv (\psi - \tilde \psi)/\sqrt{2}$, gives negligible contribution
to the total energy of the universe due to the $D$-flat condition.
The inflaton $\phi$ almost maximally mixes with $\psi_{+}$ to form the
mass eigenstates~\cite{Kawasaki:2006gs},
\beq
\label{eq:mass-eigen}
\phi_{\pm} \;\equiv\; \frac{\phi \pm \psi_{+}}{\sqrt{2}}.
\eeq
The VEVs and the masses of these mass eigenstates are given by
\beq
\label{eq:mass-vev}
\la \phi_\pm \ra \;=\; \frac{\mu}{\sqrt{\lambda}},~~~~~~
m_{\phi_\pm}\;=\; \sqrt{2} \lambda \la \phi_\pm \ra.
\eeq
Since the two mass eigenstates $\phi_\pm$ have the equal decay
rates due to the (almost) maximal mixing, we can simply treat the
reheating process just like a single-field inflation model; the
reheating is parameterized by a single parameter, $T_R$ (or
equivalently, $\Gamma_\phi$)~\footnote{
The preheating~\cite{Felder:2000hj,Asaka:2001ez,Copeland:2002ku} 
is known to occur in this model,
and if it occurs, the homogeneous mode of the inflaton and the
waterfall fields disappear soon and the excited particles are
produced. This instability itself does not affect our discussion,
since these excited particles will decay perturbatively into the
lighter fields in the end.
}. 

We have plotted the mass and the VEV given by Eq.~(\ref{eq:mass-vev})
for $\lambda = 10^{-5} - 10^{-1}$ in Figs.~\ref{fig:tr-low},
\ref{fig:1GeV}, \ref{fig:1TeV}, and \ref{fig:100TeV}. From
Fig.~\ref{fig:tr-low}, one can see that the lower bound on the
reheating temperature broadly ranges from $O(1){\rm\,GeV}$ up to
$O(10^8){\rm\,GeV}$. The gravitino problem excludes some fraction of
the parameter space; for instance, $m_{\phi_\pm} \gtrsim O(10^{13})
{\rm\,GeV}$ is excluded for $m_{3/2} = 1{\rm\,TeV}$ with $B_h = 1$,
since the reheating temperature exceeds $10^6 {\rm\,GeV}$ (see
Eq.~(\ref{eq:unstable-Y1})).  The gravitino production from the
inflaton decay actually gives severer bounds as shown in
Figs.~\ref{fig:1GeV}, \ref{fig:1TeV}, and \ref{fig:100TeV}.  In
particular, the entire parameter region is excluded for $m_{3/2} =
1{\rm\,TeV}$ with $B_h = 1$.  For much larger or smaller $m_{3/2}$,
the constraints on the hybrid inflation model is relaxed, and the
region with relatively small $m_\phi$ is allowed. Note, however, that
it then tends to be disfavored by WMAP three year
data~\cite{Spergel:2006hy} since the predicted spectral index $n_s$
approaches to unity~\footnote{ Note that the hybrid inflation produces
negligible tensor fluctuations and hence we should take
the WMAP
constraint on $n_s$ for no tensor mode, 
$n_s = 0.95 \pm 0.02$.}.  In
other words, $n_s \simeq 1$ is necessary for the hybrid inflation
model to be compatible with the gravitino overproduction from the
inflaton decay.

Next let us consider a smooth hybrid inflation
model~\cite{Lazarides:1995vr}.  The superpotential of the inflaton sector is
\beq
 \label{eq:spot_smhyb}
   W(\phi, \psi,\tilde{\psi}) = \phi \left(\mu^{2}  
   - \frac{ (\tilde{\psi}\psi)^n}{M^{2n-2}}\right),
\eeq
where $M$ is an effective cut-off scale, and $n \geq 2$ is an
integer. The vacuum of the scalar potential is located at $\la \phi
\ra = 0$ and $\la \psi \ra = \la \tilde \psi \ra=(\mu M^{n-1})^{1/n}$ 
in the global SUSY limit. Note that $\psi =
\tilde{\psi}$ always holds due to the additional $D$-term
potential. As in the hybrid inflation model, one of the combination,
$\psi^{(+)}\equiv (\psi+\tilde{\psi})/\sqrt{2}$, almost maximally
mixes with $\phi$ to form the mass eigenstates $\phi_\pm$ defined by
Eq.~(\ref{eq:mass-eigen}). The VEVs and masses of $\phi_\pm$ are given by
\beq
\label{eq:mass-vev2}
\la \phi_\pm \ra \;=\; (\mu  M^{n-1})^{1/n},~~~~~~
m_{\phi_\pm}\;=\; \sqrt{2} n \mu^2/\la \psi \ra,
\eeq
which are plotted in Figs.~\ref{fig:tr-low}, \ref{fig:1GeV},
\ref{fig:1TeV}, and \ref{fig:100TeV}, in the case of $n = 2, \,3$ and
$4$. The ranges of the parameters are determined by requiring both a
successful inflation with a large enough e-folding number and the
validity of the effective description \EQ{eq:spot_smhyb}.  The scalar
spectral index is then predicted to be $n_s \simeq 0.97$, which is
slightly smaller than the simple hybrid inflation model.  From
\FIG{fig:tr-low}, we can see that the smooth hybrid inflation model is
incompatible with the gravitino of $m_{3/2} = 1{\rm\,TeV}$, due to the
too high reheating temperature. Further, taking account of the
gravitino production from the inflaton decay, the smooth hybrid
inflation model gets in more trouble for a broad range of the
gravitino mass (see Figs.~\ref{fig:1GeV}, \ref{fig:1TeV}, and
\ref{fig:100TeV}). Note also that the constraints get severer for
larger $n$.

Lastly, let us mention that those problems stated above can be avoided
in the chaotic inflation model with a $Z_2$ symmetry.  A chaotic
inflation~\cite{Linde:1983gd} is realized in SUGRA, based on a
Nambu-Goldstone-like shift symmetry of the inflaton chiral multiplet
$\phi$~\cite{Kawasaki:2000yn}. Namely, we assume that the K\"ahler
potential $K(\phi,\phi^\dag)$ is invariant under the shift of $\phi$,
\begin{equation}
  \phi \rightarrow \phi + i\,A,
  \label{eq:shift}
\end{equation}
where $A$ is a dimensionless real parameter. Thus, the K\"ahler
potential is a function of $\phi + \phi^\dag$; $K(\phi,\phi^\dag) =
K(\phi+\phi^\dag)= c\,(\phi+\phi^\dag) + \frac{1}{2}
(\phi+\phi^\dag)^2 + \cdots$, where $c$ is a real constant and must be
smaller than $O(1)$ for a successful inflation.  We will identify its
imaginary part with the inflaton field $\varphi \equiv \sqrt{2} {\rm
\,Im}[\phi]$.  Moreover, we introduce a small breaking term of the
shift symmetry in the superpotential in order for the inflaton
$\varphi$ to have a potential:
\begin{equation}
  W(\phi,\psi) = m\,\phi \,\psi, 
  \label{eq:mass}
\end{equation}
where we introduced a new chiral multiplet $\psi$, and $m \simeq
10^{13}$GeV determines the inflaton mass.  The scalar potential is
given by
\beq
V(\eta,\varphi,\psi) \;\simeq\; \frac{1}{2}m^2 \varphi^2 + m^2 |\psi|^2,
\eeq
where we set the real part of $\phi$ to be at the vacuum.  For
$\varphi \gg 1$ and $|\psi| < 1$, the $\varphi$ field dominates the
potential and the chaotic inflation takes place (for details see
Refs~\cite{Kawasaki:2000yn}).

Although $\varphi$ does not acquire any finite VEV, the linear term in
the K\"ahler potential behaves exactly the same as a VEV. Therefore,
the decay rates given by Eqs.~(\ref{eq:rate-th-top}) and
(\ref{gravitino-rate}) apply to the inflaton $\varphi$, if one
replaces $\la \phi \ra$ with $c$, the coefficient of the linear term.
If $c$ is sizable, the chaotic inflation model of this type as well
may encounter the cosmological difficulties. However, one can suppress
such linear term by assuming an approximate $Z_2$ symmetry. Therefore
the problems mentioned above can be avoided in the chaotic inflation
model.

\section{Discussion and Conclusions}
\label{sec:4}

So far we have focused on the inflaton decay into the visible sector
and the gravitinos. Since the inflaton couples to all matter fields
once it acquires a finite VEV, the inflaton can also decay into the
hidden and/or messenger sector. The decay may cause another
cosmological problem. For instance, if the messenger fields are
produced by the inflaton decay, and if the lightest messenger particle
is stable, the abundance of such stable particle may easily exceed the
present DM abundance.  The production of the hidden sector field may
be faced with the similar problem. Therefore, the inflaton decay
process shown in this letter can put a constraint on the structure of
the hidden and/or messenger sector.

The SUSY breaking may occur at tree level as well, although we have
assumed the DSB scenarios in the previous sections.  Our discussion on
the gravitino production from the inflaton decay actually applies to
any SUSY breaking models containing Yukawa interactions with sizable
couplings. For instance, the O'Raifeartaigh-type
models~\cite{O'Raifeartaigh:1975pr} may contain such terms.  Note that
a linear term in the superpotential as in the Polonyi model only
induces a small mixing with the inflaton.

It depends on the Yukawa coupling constants in the SUSY breaking
sector how much the gravitinos are produced by the inflaton decay.
Although we have assumed $C \gtrsim O(1)$ in the above discussion, $C$
may be smaller if all the couplings are extremely suppressed.  If this
is the case, the constraints shown in Figs.~\ref{fig:1GeV},
\ref{fig:1TeV} and \ref{fig:100TeV} are relaxed. Note that, in this
case, the gravitino production sets severe constraint on the SUSY
breaking sector, instead of the inflaton parameters.  We have assumed
no entropy production late after the reheating of inflation throughout
this letter. If a late-time entropy production
occurs~\cite{Lyth:1995ka}, the constraints derived in the previous
sections can be relaxed. Another even manifest solution to the
gravitino overproduction problem is to assume the gravitino mass
$m_{3/2} < O(10)\,$eV~\cite{Viel:2005qj}.  In this case, the produced
gravitinos get into thermal equilibrium due to relatively strong
interactions with the standard-model particles, and such light
gravitinos are cosmologically harmless.

Since the inflaton with a nonzero VEV couples to all matter fields
that appear in the superpotential, it also decays into the
right-handed neutrinos through the large Majorana mass term. The
non-thermally produced right-handed neutrinos may generate the baryon
asymmetry of the universe through the
leptogenesis~\cite{Fukugita:1990gb, LGinf}.  Since the abundance of
the right-handed neutrino is generically correlated with that of the
gravitino produced in a similar way, $C \ll 1$ is required for the
unstable gravitinos to realize the successful leptogenesis. However,
for the stable gravitinos, the non-thermal leptogenesis may work even
if $C = O(1)$. This opens up an interesting way to induce the
non-thermal leptogenesis; the right-handed neutrinos are produced by
the inflaton decay even if there is no direct coupling in the global
SUSY limit; they are produced simply because the Majorana mass is
large (but smaller than the inflaton mass).  Detailed discussion on
this topic will be presented elsewhere.

In this letter, we have shown that, once the inflaton acquires a
finite VEV, it can decay into all matter fields via the SUGRA effect,
as long as kinematically allowed. It is a striking feature that the
decay occurs even without the direct couplings in the global SUSY
limit.  The inflaton with a nonzero VEV can therefore decay into the
visible sector fields through the top Yukawa coupling, which has
enabled us to set a lower limit on the reheating temperature.  For a
typical inflaton mass in high-scale inflations such as the hybrid and
chaotic inflation models, the inflaton can decay into the SUSY
breaking sector fields, producing the gravitinos in a broad class of
the DSB models. We have seen that the gravitino production from the
inflaton decay severely constrains the high-scale inflation models. We
would like to stress again that the gravitino production from the
inflaton decay is a generic problem; it is present even in the gauge-
and anomaly-mediation models, since the decay proceeds irrespective of
whether the SUSY breaking field $z$ is charged under some symmetries
or not.  In particular, the gravitinos are produced even if the
K\"ahler potential is minimal.  One of the solution is to assign a
symmetry on the inflaton field to forbid a nonzero VEV and a linear
term in the K\"ahler potential. In fact, the chaotic inflation model
with an approximate $Z_2$ symmetry can avoid the problem.

\section*{Acknowledgments}
M.E and F.T.  would like to thank the Japan Society for Promotion of
Science for financial support.  The work of T.T.Y. has been supported
in part by a Humboldt Research Award.


\clearpage

\end{document}